\documentstyle[sprocl]{article}

\def\be{\begin{equation}}
\def\ee{\end{equation}}
\def\bea{\begin{eqnarray}}
\def\eea{\end{eqnarray}}

\begin{document}
\begin{flushright}
{\bf
TUW 00-03 \\
hep-th/0001149 \\
2000, January 20}
\end{flushright}

\title{
ON ACTION FUNCTIONALS FOR INTERACTING BRANE SYSTEMS}

\author{I. BANDOS},

\address{Institute for Theoretical Physics, \\
NSC Kharkov Institute of Physics and Technology,\\
 310108, Kharkov,  Ukraine \\
e-mail: bandos@hep.itp.tuwien.ac.at \\
bandos@kipt.kharkov.ua}

\author{W. KUMMER}

\address{Institut f\"{u}r Theoretische Physik,
\\
Technische
 Universit\"{a}t Wien,
\\ Wiedner Hauptstrasse 8-10, A-1040 Wien\\
 e-mail: wkummer@tph.tuwien.ac.at}

\maketitle\abstracts{
We present an action functional and derive equations of motion for a
coupled system of a bosonic Dp--brane and an open string ending on the
Dp-brane.  With this example we address the key issues of the recently
proposed method \cite{BK,BKD9} for the  construction of  manifestly
supersymmetric action functionals for interacting superbrane systems.  We
clarify, in particular,  how the arbitrariness in sources localized on the
intersection \cite{BK,BKD9}  is related to the standard description of
the flat D-branes as rigid planes where the string for endpoints 'live'.
}

\section*{Introduction}

Recently an approach to obtain a supersymmetric action functional for a
coupled system of interacting superbranes has been proposed
\cite{BK,BKD9}.
The aim of the study \cite{BK,BKD9} was to construct a basis for a future
search for new solitonic solutions  of worldvolume equations of superbranes
and of supergravity equations with a complicated system of brane sources.

The system of open fundamental superstring ending on a super--Dp-brane
\cite{Dbranes} was considered in \cite{BK,BKD9} as a generic case of
interacting brane system in the frame of the 'brane democracy' concept
\cite{Town}.  On the other hand, the manifestly supersymmetric
description of such system can be useful in a search for a quasiclassical
effective action of the system of coincident branes which are conjectured
to carry non-Abelian gauge fields \cite{NDp} (non-Abelian
Dirac--Born--Infeld action \cite{Tseytlin}).

In this contribution we would like to discuss the key issues
of  our approach \cite{BK,BKD9} by studying a pure bosonic limit of the
interacting brane system. We elaborate the case of open bosonic string
with endpoints living on the {\sl dynamical} bosonic  Dp--brane and
discuss the relation of the approach \cite{BK,BKD9} with the standard
description \cite{Dbranes}
where the Dp-branes are treated as rigid
$p$-dimensional hyperplanes.

\section{Standard description of open string}

To clarify the problem and to establish the notations let us begin with the
standard consideration of the open bosonic string in flat
D--dimensional space--time ${\underline{{\cal M}}^{D}}$.
The action can be written as \cite{gsw}
\begin{equation}\label{Sbstr}
S_{str}=  {1 \over 2} \int_{{\cal M}^2} d^2 \xi \sqrt{-g} g^{mn}(\xi )
\partial_m \hat{X}^{\underline{m}}(\xi ) \partial_n
\hat{X}^{\underline{n}}(\xi ) \eta_{\underline{m} \underline{n}}.
\end{equation}
Here $\{ (\xi^m)\}= \{ (\tau,\sigma )\}$ ($m=0,1$) are local coordinates on the string worldsheet ${\cal M}^2$, $\hat{X}^{\underline{m}}(\xi )$ (
$\underline{m}=0,\ldots , (D-1) $) are coordinate (embedding) functions
which define (locally) an embedding of the worldsheet  ${\cal M}^2$ into the
D--dimensional space--time (target) ${\underline{{\cal M}}^{D}}$
 \begin{equation}\label{bembf}
{X}^{\underline{m}}= \hat{X}^{\underline{m}}(\xi ):
\qquad {\cal M}^2 \quad \rightarrow \underline{{\cal M}}^{D}.
\end{equation}
The flat metric of ${\underline{{\cal M}}^{D}}$ is chosen to be 'mostly minus'
$\eta_{\underline{m} \underline{n}}= diag(+,-,\ldots,-)$.

When the worldsheet has a boundary $\partial {\cal M}^2$, the
variation of the action (\ref{Sbstr}) acquires the form
($d\xi^m \wedge d\xi^n = \varepsilon^{mn} d^2 \xi$,
$(d\xi)^{\wedge 2} = 1/2 \varepsilon_{mn} d\xi^m \wedge d\xi^n =
- d^2 \xi$)
$$
\delta  S_{str} =
{1 \over 2} \int_{{\cal M}^2} d^2 \xi \sqrt{-g} \delta g^{mn}(\xi )
\left(
\partial_{m} {\hat X}^{\underline{m}}
\partial_{n} {\hat X}_{\underline{m}}
- {1 \over 2} g_{mn} g^{kl}\partial_{k} {\hat X}^{\underline{m}}
\partial_{l} {\hat X}_{\underline{m}}
\right)
$$
\begin{equation}\label{vSbstr}
- \int_{{\cal M}^2} d^2 \xi
\delta {\hat X}_{\underline{m}}\partial_{m} \Big( \sqrt{-g} g^{mn}
\partial_{n} {\hat X}^{\underline{m}}\Big)
- \int_{\partial {\cal M}^2}
d \xi^m \varepsilon_{mn}
\sqrt{-g} g^{nk} \partial_{k} {\hat X}^{\underline{m}}
\delta {\hat X}_{\underline{m}}.
\end{equation}
%$$
%$$

The variation with respect to the auxiliary intrinsic metric
$g_{mn}(\xi )$ results in the equation
\begin{equation}\label{vg}
\partial_m \hat{X}^{\underline{m}}(\xi )
\partial_n \hat{X}_{\underline{n}}(\xi ) = {1 \over 2}
g_{mn}(\xi )
g^{pq}(\xi )
\partial_p \hat{X}^{\underline{m}}(\xi )
\partial_q \hat{X}_{\underline{n}}(\xi ).
\end{equation}
One easily finds that the trace part of Eq. (\ref{vg})  is satisfied
identically. This is the Noether identity reflecting the
gauge Weyl symmetry of the action (\ref{Sbstr})
\begin{equation}\label{Wg}
g_{mn}(\xi) \rightarrow g^\prime_{mn}(\xi) = e^\Lambda g_{mn}(\xi) ,
\qquad
X^{\underline{m}}(\xi) \rightarrow
X^{\underline{m}}(\xi) \; .
\end{equation}
With this symmetry one can fix the gauge
\begin{equation}\label{Wgg}
g^{mn}(\xi ) \partial_m \hat{X}^{\underline{m}}(\xi )
\partial_m \hat{X}_{\underline{n}}(\xi )=2.
\end{equation}
Then Eq. (\ref{vg}) becomes
\begin{equation}\label{imstr}
g_{mn}(\xi ) = \partial_m \hat{X}^{\underline{m}}(\xi )
\partial_m \hat{X}_{\underline{n}}(\xi ),
\end{equation}
and implies that the metric is induced by the embedding.

The {\sl usual way} to deal with the coordinate variation
of the action (\ref{Sbstr}) (presented in the second line
of Eq. (\ref{vSbstr}))
is to assume that the {\sl bulk and the boundary inputs
 should vanish separately}.
Then one arrives at the well known result that
\begin{itemize}
\item proper equations of motion
have the form
\begin{equation}\label{emstr}
\partial_m \left( \sqrt{-g} g^{mn}(\xi )
\partial_n \hat{X}^{\underline{m}}(\xi ) \right)=0;
\end{equation}
\item the boundary conditions should be chosen in a way which provides
\begin{equation}\label{bkstr}
d \xi^m \varepsilon_{mn} \sqrt{-g} g^{nk}
\partial_k \hat{X}^{\underline{m}} \delta \hat{X}_{\underline{m}}
\vert_{\partial {\cal M}^2 : ~~\xi^m = {\tilde \xi}^m(\tau) } = 0
\end{equation}
where the functions ${\tilde \xi}^m(\tau)$ define parametrically the
'embedding'
of (a connected piece of) the world sheet boundary $\partial {\cal M}^2$
into the worldsheet
%${\cal M}^2$
 \begin{equation}\label{bembb}
\xi^m = {\tilde \xi}^m (\tau) :
\qquad
\partial {\cal M}^2 \quad \rightarrow {{\cal M}}^{2}.
\end{equation}
\end{itemize}

To satisfy Eq. (\ref{bkstr}) in a simple way one can
assume that for some part ${\hat X}^i(\xi)$ of the embedding functions
 \begin{equation}\label{split}
{\hat X}^{\underline{m}}= ({\hat X}^{a}, {\hat X}^{i}), \quad a=0,\ldots p,
\quad i= (p+1), \ldots , (D-1)
\end{equation}
the Neumann boundary conditions are imposed
\begin{equation}\label{Nbc}
d \xi^m \varepsilon_{mn} \sqrt{-g} g^{nk}
\partial_{k} {\hat X}^{a}
\vert_{\xi^m = {\tilde \xi}^m(\tau)} = 0, \quad \Leftrightarrow \quad
\partial_{\perp }  {\hat X}^{a} \vert_{\partial {\cal M}^2} = 0,
\end{equation}
while the remaining equations from the set  (\ref{bkstr}) are satisfied
due to the Dirichlet boundary conditions. The latter can be treated as
ones imposed {\sl on the variation} $\delta {\hat X}^{i} ({\tilde
\xi}^m(\tau)) = 0$ or, equivalently, as the boundary conditions imposed on
the coordinate functions $\hat{X}^i(\xi )$ directly
\begin{equation}\label{Dbc}
{\hat X}^{i} ({\tilde \xi}^m(\tau))= {\tilde
X}^i (= const) \qquad \Leftrightarrow
\qquad
\delta {\hat X}^{i} ({\tilde
\xi}^m(\tau)) = 0
\end{equation}

Note that {\sl for the Dirichlet boundary problem the equations
(\ref{emstr}) appear formally only on the proper open subset of the
worldsheet, i.e.\  only outside the boundary}, because the variation on the
boundary vanishes just due to Eq. (\ref{Dbc}).  In the Neumann problem we
can regard Eq. (\ref{emstr}) as valid on the whole worldsheet. But in this
case we shall assume that the boundary conditions (\ref{Nbc}) {\sl are
imposed after the variation of the action has been performed} (as we did
not restrict the variations by the condition (\ref{Nbc}) when derived the
equations of motion).

The boundary conditions (\ref{Nbc}), (\ref{Dbc})
certainly break Lorentz invariance for $p \not= (D-1)$.
Such a breaking reflects the existence of some $d=(p+1)$-dimensional
defects  where the string endpoints move. These are
worldvolumes of
Dp-branes \cite{Dbranes} considered as flat
hyperplanes.

\section{Extended variational problem}

However, we can proceed in a different manner which we will call
{\sl extended variational problem} or {\sl extended variational approach}.

Let us introduce the current density distribution with support on the
boundary of the worldsheet ($d\xi^{m}\wedge d\xi^{n}= d^2\xi \epsilon^{mn}$)
\begin{equation}\label{j1}
j_1 = d\xi^{m} \varepsilon_{mn} \int_{\partial {\cal M}^{2}}
d{\tilde \xi}^{n} (\tau )
\delta^{2} \left( \xi - {\tilde \xi} (\tau )\right)~=
\varepsilon_{mn} d\xi^{m} j^{n}, \qquad
\end{equation}
Then one easily find that
\begin{equation}\label{j1A}
\int_{{\cal M}^{2}} j_1 \wedge
{\hat {\cal A}}_1  =
\int_{\partial {\cal M}^{2}} {\tilde{\hat {\cal A}}}_1
\end{equation}
holds for any one-form ${\hat{\cal A}}_1= d\xi^m {\cal A}_m (\xi)
$ defined on the worldsheet, $
{\tilde{\hat {\cal A}}}_1= d\tilde{\xi}^m (\tau) {\cal A}_m
(\tilde{\xi}(\tau))$.

With the use of $j_1$ the coordinate variation of the string action
can be written in the form
%$$
%\delta  S_{str} =
%{1 \over 2} \int_{{\cal M}^2} d^2 \xi
%\sqrt{-g} \delta g^{mn}(\xi ) \left(
%\partial_{m} {\hat X}^{\underline{m}}(\xi )
%\partial_{n} {\hat X}_{\underline{m}}(\xi ) - g_{mn} \right)
%$$
\be\label{v2}
\delta  S_{str} =
- \int_{{\cal M}^2} \left( d^2 \xi
\partial_{m} \left( \sqrt{-g} g^{mn}
\partial_{n} {\hat X}^{\underline{m}} \right)
- j_1 \wedge d \xi^m \varepsilon_{mn}
\sqrt{-g} g^{nk} \partial_{k} {\hat X}^{\underline{m}} \right)
\delta {\hat X}_{\underline{m}}
\ee

The equations of motion are (\ref{imstr}) and
\be\label{emj}
d^2\xi \partial_{m} \left( \sqrt{-g} g^{mn}(\xi )
\partial_{n} {\hat X}^{\underline{m}}(\xi ) \right)  =
j_1 \wedge d \xi^m \varepsilon_{mn}
\sqrt{-g} g^{nk} \partial_{k} {\hat X}^{\underline{m}}
\ee
$$
\Leftrightarrow \quad \partial_{m} \left( \sqrt{-g} g^{mn}
\partial_{n} {\hat X}^{\underline{m}}\right)  =
- \int_{\partial{\cal M}^2} d\tilde{\xi}^m \varepsilon_{mn}
\sqrt{-g} g^{nk} \partial_{k} {\hat X}^{\underline{m}}
\delta^2 \left( \xi - \tilde{\xi}(\tau)\right).
$$

Thus no boundary conditions appear, but the equations of motion acquire
a source localized at the boundary of the worldsheet.
The result (\ref{emj}) of such an {\sl 'extended variational problem'}
looks
%apparently
different from the
standard one (\ref{emstr}), (\ref{Nbc}), (\ref{Dbc}).
% and thus from the
%standard statement of problem characteristic for  Classical Mathematical
%Physics.

%The reason is that we actually extended the space of admissible functions
%in the statement of variational problem (\ref{v2}) including distributions
%into the consideration. So one can call the approach (\ref{v2})
%{\sl extended variational problem}.
%
%A question arises: whether we can reproduce the classical statement of
%problem
%with free equations (\ref{emstr}) and boundary conditions (\ref{Nbc}),
%(\ref{Dbc}) at least as a particular solution of the extended
%variational problem?

To clarify the situation
let us, for the moment, fix the  conformal gauge
and
%, for definiteness,
accept the parametrization where
the string endpoints correspond to the values
$\sigma =0$ and $\sigma =\pi$.
Then  the standard approach produces the  free (Laplace) equations with
boundary conditions
\begin{equation}\label{Ncg}
(\partial^2_\tau -  \partial^2_\sigma ) \hat{X}^a (\tau,\sigma ) = 0, \qquad
\partial_\sigma \hat{X}^a \vert_{{\cal M}^2}=0,
\end{equation}
\begin{equation}\label{Dcg}
(\partial^2_\tau -  \partial^2_\sigma ) \hat{X}^i (\tau,\sigma ) = 0, \qquad
\hat{X}^i (\tau , 0) = \tilde{X}^i_0,   \quad \hat{X}^i (\tau , \pi ) =
\tilde{X}^i_\pi ,
\end{equation}
while the extended variational problem results in the equations
with sources for both types of the coordinate functions
\begin{equation}\label{a1}
(\partial^2_\tau -  \partial^2_\sigma ) \hat{X}^a (\tau,\sigma ) =
\partial_\sigma  \hat{X}^a \left(\delta (\sigma)- \delta (\sigma - \pi)
\right)
\end{equation}
\begin{equation}\label{i1}
(\partial^2_\tau -  \partial^2_\sigma ) \hat{X}^i (\tau,\sigma ) =
\partial_\sigma  \hat{X}^i \left(\delta (\sigma)- \delta (\sigma - \pi )\right)
\end{equation}
It is evident that the problem (\ref{Ncg}) can be obtained
as a particular
case of (\ref{a1}) because, after imposing an additional
condition $\partial_\sigma \hat{X}^a \vert_{{\cal M}^2}=0$, the
r.h.s. of Eq. (\ref{a1}) vanishes. As in such a way we arrive at the free
equations on the whole worldsheet, this corresponds just to the result of
Neumann variation problem with the supposition that the boundary
conditions are imposed {\sl after} the variation has been performed (i.e.
imposed 'by hand').

This does {\sl not} hold for
(\ref{Dcg}), because $\hat{X}^i \vert_{{\cal M}^2}=const $ does not imply
$\partial_\sigma \hat{X}^i \vert_{{\cal M}^2}=0$.
Thus the value of Laplace operator acting on the coordinate
function {\sl is not indefinite} on the boundary, as it is in the classical
Dirichlet problem, but is determined by the value of the derivative
$\partial_\sigma \hat{X}^i \vert_{{\cal M}^2}$.

The fact that the straightforward application of the
extended variational approach does not produce the Dirichlet
boundary problem even as a particular case does not look not so surprising
if one remembers that it assumes that the variations of the coordinate functions
are unrestricted everywhere, while the Dirichlet boundary conditions
(\ref{Dbc}) come just as restrictions on the variations.

\section{Lagrange multiplier approach for Dirichlet problem.}

If one would like to reproduce the
Dirichlet problem (\ref{Dcg}) from the extended variational approach,
one can try to incorporate the boundary conditions (\ref{Dbc})
with Lagrange multiplier one-form
$P_1^i\equiv d\tau {\cal P}^i(\tau )$
into the action written  in the conformal gauge
$g_{mn}= \partial_m X^{\underline{m}}\partial_n X_{\underline{m}}
= \eta_{mn} \equiv diag (+1,-1)$ (cf. \cite{BK})
\begin{equation}\label{Sbstrc1}
S_{str}=  {1 \over 2} \int_{{\cal M}^2} d^2 \xi
\left( \partial_m \hat{X}^a \partial^m \hat{X}_a -
\partial_m \hat{X}^i \partial^m \hat{X}^i  \right)
+
 \int_{\partial {\cal M}^2} d \tau
{\cal P}_i \left( \hat{X}^i \left(\xi(\tau)\right)-
\tilde{X}^i  \right)
\end{equation}
Then the Dirichlet boundary conditions are produced by the
variation with respect to Lagrange multiplier
 ${\cal P}^i (\tau)$ or, more precisely, two Lagrange multipliers
${\cal P}_i^0 (\tau )$ and ${\cal P}_i^\pi (\tau )$ (each living on the
worldline of the corresponding endpoint of the string),
while the dynamical equations
for the coordinate functions $\hat{X}^i$ read
\begin{equation}\label{i2}
(\partial^2_\tau -  \partial^2_\sigma ) \hat{X}^i (\tau,\sigma ) =
\left(\partial_\sigma  \hat{X}^i  - {\cal P}_i^0 \right)\delta (\sigma)
- \left(\partial_\sigma  \hat{X}^i  - {\cal P}_i^\pi \right)
\delta (\sigma - \pi ).
\end{equation}
What one can see from the very beginning is the arbitrariness
in the Lagrangian multipliers
${\cal P}_i^{0, \pi} (\tau )$
which cannot be fixed by any gauge symmetry
of  the action.
So one can conclude that  the Lagrange multipliers     carry some
degrees of freedom
and some doubts may arise concerning the applicability of the
Lagrangian multiplier method to the problem.

However if one notes that\\
i) Eq. (\ref{i2}) becomes free when considered  outside the boundary
(i.e. on the open proper subset of the string worldsheet), \\
ii) the boundary conditions (\ref{Dbc}) are produced as equations
of motion for Lagrange multipliers,
\\
iii) the action of the
Laplace operator on the coordinate functions $X^i$
is {\sl indefinite}  on the boundary
just due to the {\sl arbitrariness}
of the Lagrange multiplier, \\
one easily concludes that the Dirichlet boundary problem is
indeed reproduced.

%Substituting
When one can find
a solution for it
(as it can be done in the case of string), then, substituting it into the
equations, one can fix the value of the Lagrange multiplier ${\cal
P}^i (\tau )$.

This way to reproduce the Dirichlet boundary problem could be regarded
as an artificial one. However it just provides the possibility
to describe the system of open (super)string and {\sl dynamical}
(super-)Dp--brane and, more generally, of open (super--)brane(s)
ending on closed 'host' (super--)brane(s)
on the level of (quasi)classical action functional \cite{BK}.

\section{Dynamical string with the ends on dynamical D-brane}

The action for free Dp-brane \cite{Dbranes,c1} has the form
\be\label{Dp}
S_{Dp} = \int_{{\cal M}^{p+1}} d^{p+1} \zeta
\sqrt{|G|}, \qquad |G| = (-)^p det\left(G_{\tilde{m}\tilde{n}}\right) ,
\qquad
\ee
where
\be\label{GgF}
G_{\tilde{m}\tilde{n}}\equiv
\partial_{\tilde{m}} \tilde{X}^{\underline{m}}(\zeta )
\partial_{\tilde{n}} \tilde{X}_{\underline{m}}(\zeta )
- {\cal F}_{\tilde{m}\tilde{n}}
\ee
is the non-symmetric  'open string induced metric' \cite{S&W},
the coordinate functions $\hat{X}^{\underline{m}} (\zeta^{\tilde{m}})$
define (locally) an embedding of the $d=(p+1)$-dimensional worldvolume
${\cal M}^{p+1}= \{ (\zeta^{\tilde{m}} ) \}$ ($\tilde{m}=0,\ldots, p$)
of the Dp-brane into the
D(=10)--dimensional space--time  ${\underline{{\cal M}}^{D}}$
 \begin{equation}\label{bembD}
{X}^{\underline{m}}= \tilde{X}^{\underline{m}}(\zeta ):
\qquad {\cal M}^{p+1} \quad \Rightarrow \quad \underline{{\cal M}}^{D},
\end{equation}
 \begin{equation}\label{FdA}
{\cal F}\equiv dA = {1 \over 2} d\zeta^{\tilde{m}}
d\zeta^{\tilde{n}} {\cal F}_{\tilde{n}\tilde{m}}
\end{equation}
 is the field strength of the
worldvolume gauge field of the Dp--brane
$A= d\zeta^{\tilde{m}} A_{\tilde{m}}( \zeta )$.

With this notation the variation of the {\sl free}
Dp-brane action (\ref{Dp}) can be
written in the form
\be\label{vDp}
\delta S_{Dp}=
+ \int_{{\cal M}^{p+1}} d^{p+1} \zeta \partial_{\tilde{m}} \left(
\sqrt{|G|} G^{-1 [\tilde{m}\tilde{n}] }\right) \delta A_{\tilde{n}}
(\zeta ) \ee
$$ - \int_{{\cal M}^{p+1}} d^{p+1} \zeta \partial_{\tilde{m}}\left(
\sqrt{|G|}
~{G}^{-1 (\tilde{m}\tilde{n})}
\partial_{\tilde n}\tilde{X}^{\underline{m}} \right)
\delta  \tilde{X}_{\underline{m}},
$$
where
$$ G^{-1~mn} = G^{-1 (mn)}   + G^{-1 [mn]}  $$
is the matrix inverse to the open string metrics (\ref{GgF}).

To describe the interacting system of the open string and the {\sl
dynamical} D-brane on the level of action functional one has to provide
the counterpart of (\ref{Dbc}) with {\sl coordinate functions of D-brane}
instead of constants $\tilde{X}^i$. Namely we have to impose an
identification \be\label{id}
\hat{X}^{\underline{m}} \left(\tilde{\xi}(\tau )\right) =
\tilde{X}^{\underline{m}} \left(\hat{\zeta }(\tau )\right),
 \ee
where $\zeta^{\tilde{m}}=\hat{\zeta }^{\tilde{m}}(\tau )$
defines an embedding of the worldline(s) of string endpoints
(i.e. the boundary of the worldsheet) into the worldvolume of the
Dp-brane
\be\label{bD}
\zeta^{\tilde{m}}=\hat{\zeta }^{\tilde{m}}(\tau ) : \qquad
\partial {\cal M}^{2} \quad \Rightarrow \quad {\cal M}^{p+1}.
\ee

The problem is that, when (\ref{id}) is taken into account
the variations $\delta \hat{X}^{\underline{m}} $ and
$\delta\tilde{X}^{\underline{m}} \vert_{\zeta=\hat{\zeta }(\tau )}$
cannot be treated as independent on the intersection
${\cal M}^{1+p} \cup {\cal M}^{2}= \partial {\cal M}^2$.
Indeed, varying (\ref{id}) one finds
\be\label{vid}
\delta \hat{X}^{\underline{m}} \vert_{\xi= \tilde{\xi}(\tau )}
+ \delta \tilde{\xi}^m (\tau )
\partial_m\hat{X}^{\underline{m}}
\left( \tilde{\xi}(\tau )\right) =
\delta \tilde{X}^{\underline{m}} \vert_{\zeta=\hat{\zeta }(\tau )}+
\delta \hat{\zeta}^{\tilde{m}}(\tau )
\partial_{\tilde{m}} \tilde{X}^{\underline{m}}
\left(\hat{\zeta }(\tau )\right).
 \ee

A method  to implement (\ref{id}) into the action principle was
proposed in \cite{BK}.
It consists in imposing  the  identification (\ref{id}) with
Lagrange multipliers
one-form $P_{1}^{\underline{m}}= d\tau {\cal P}^{\underline{m}}(\tau)$
into the action (cf. with (\ref{Sbstrc1})).
Thus the  {\sl action for a coupled system} of open string and
{\sl dynamical} Dp-brane can be written as
\be\label{Sg}
S= S_{str} + S_{Dp} + S_{int}.
\ee
Here
$S_{str}$ is determined by (\ref{Sbstr}) but with
{\sl open} worldsheet whose boundary is a set of two (or more!) worldlines
which are embedded into the worldvolume of the Dp--brane (\ref{bD}).
$S_{Dp}$ is determined by Eq. (\ref{Dp}). The last term in
(\ref{Sg})
\be\label{Sint} S_{int} = \int_{\partial {\cal M}^{2}} \left(\hat{A} +
P_{1\underline{m}} \left( \hat{X}^{\underline{m}}(\tilde{\xi } (\tau )) -
\tilde{X}^{\underline{m}} (\hat{\zeta} (\tau )) \right) \right).
\ee
describes, in particular, the interaction of the string endpoints
with the worldvolume gauge field of the Dp--brane $A=
d\zeta^{\tilde{m}} A_{\tilde{m}}( \zeta )$ whose pull-back on the
intersection is $\hat{A}= d\hat{\zeta}^{\tilde{m}}(\tau) A_{\tilde{m}}
(\hat{\zeta}(\tau)) =d\tau \partial_\tau \hat{\zeta}^{\tilde{m}}
A_{\tilde{m}}(\hat{\zeta})$.
%The second term in (\ref{Sint})
%involves the boundary conditions on the worldvolume of the
%dynamical Dp-brane incorporated with Lagrange multiplier 1--form

The variation of the interaction term is
\be\label{vSint}
\delta S_{int}=
 \int_{\partial {\cal M}^{2}} \delta
  P_{1\underline{m}} \left[\hat{X}^{\underline{m}}(\tilde{\xi } (\tau )) -
\tilde{X}^{\underline{m}} (\hat{\zeta} (\tau ) )\right] \; +
\ee
$$
+ \int_{\partial {\cal M}^{2}}
  P_{1\underline{m}}
\left[ \delta \hat{X}^{\underline{m}}(\tilde{\xi } (\tau )) -
\delta \tilde{X}^{\underline{m}} (\hat{\zeta} (\tau ) )\right]
+ \int_{\partial {\cal M}^{2}} d\hat{\zeta}^{\tilde n} (\tau )
\delta A_{\tilde n}\left( \hat{\zeta} (\tau )\right) +
$$
$$
+
\int_{\partial {\cal M}^{2}} \delta \hat{\zeta}^{\tilde{m}} (\tau )
\left[ i_{\tilde{m}}{\cal F} - P_{1\underline{m}}\partial_{\tilde{m}}
\tilde{X}^{\underline{m}}\right] +
$$
$$
+ \int_{\partial {\cal M}^{2}} \delta \hat{\xi}^{{m}} (\tau )
\left[ P_{1\underline{m}} \partial_{{m}}
\hat{X}^{\underline{m}} -
d \xi^p \varepsilon_{pn}
\sqrt{-g} g^{nk} \partial_{k} {\hat X}^{\underline{m}}
\partial_m {\hat X}_{\underline{m}}
\right].
$$
The third and forth lines
include the variation  with respect to
coordinate functions $\hat{\zeta}^{\tilde{m}} (\tau )$
and $\tilde{\xi}^m (\tau )$
defining the embedding of the string boundary worldlines in the
Dp--brane worldvolume (\ref{bD}) and in the worldsheet
(\ref{bembb}).

\bigskip

To vary the whole action it is convenient to introduce
the following current density
distribution form (see \cite{BK} and refs. therein)\footnote{
In our conventions
$d{\zeta}^{\tilde{m}_1} \wedge \ldots  \wedge d{\zeta}^{\tilde{m}_{p+1}}=$
$\varepsilon^{\tilde{m}_1 \ldots  \tilde{m}_{p+1}}
(d\zeta)^{\wedge (p+1)} \equiv$ $ (-)^p
\varepsilon^{\tilde{m}_1 \ldots  \tilde{m}_{p+1}}
d^{p+1}\zeta $,
$d{\zeta}^{\wedge p}_{\tilde{m}_1}\equiv
\varepsilon_{\tilde{m}_1 \tilde{m}_2\ldots  \tilde{m}_{p+1}}
d{\zeta}^{\tilde{m}_2} \wedge \ldots  \wedge d{\zeta}^{\tilde{m}_{p+1}}$,
$\ldots $
}
\begin{equation}\label{j30}
j_{p} = d\zeta^{\wedge p}_{\tilde{m}} \int_{\partial {\cal M}^{p+1}}
d{\hat \zeta}^{\tilde{m}} (\tau )
\delta^{p+1} \left( \zeta - {\hat \zeta} (\tau )\right)~=
(-)^p d\zeta^{\wedge p}_{\tilde{n}} j^{\tilde{n}} . \qquad
\end{equation}
As $j_p \wedge d{\zeta}^{\tilde{m}}= d^{p+1}\zeta ~j^{\tilde{m}}$,
it is easy to see that
\begin{equation}\label{j31}
 \int_{{\cal M}^{1+3}} j_3 \wedge {\tilde {\cal A}}_1 =
\int_{\partial {\cal M}^{1+1}} {\hat {\cal A}}_1 .
\end{equation}

With the use of (\ref{j30}) and the worldsheet current density
distribution (\ref{j1}) one can write
 the variation of the action (\ref{Sg}) in the form
\\ ($ G_{\tilde{m}\tilde{n}}\equiv
\partial_{\tilde{m}} \tilde{X}^{\underline{m}}(\zeta )
\partial_{\tilde{n}} \tilde{X}_{\underline{m}}(\zeta )
- {\cal F}_{\tilde{m}\tilde{n}}$ (\ref{GgF}), $G^{-1 \tilde{m}\tilde{q}}
G_{\tilde{q}\tilde{n}}=\delta^{\tilde{m}}_{~\tilde{n}}$)

\begin{equation}\label{vSg}
\delta S = {1 \over 2} \int_{{\cal M}^2} d^2 \xi \sqrt{-g} \delta g^{mn}(\xi )
\left(\partial_{m} {\hat X}^{\underline{m}}
\partial_{n} {\hat X}_{\underline{m}} -
 {1 \over 2} g_{mn} g^{kl}\partial_{k} {\hat X}^{\underline{m}}
\partial_{l} {\hat X}_{\underline{m}}  \right) -
\end{equation}
$$
 - \int_{{\cal M}^2}
\delta {\hat X}_{\underline{m}}
\left[ d^2 \xi \partial_{m} \Big( \sqrt{-g} g^{mn} \partial_{n}
{\hat X}^{\underline{m}} \Big)
- j_1 \wedge  \left( P_1^{\underline{m}} +
d \xi^m \varepsilon_{mn}
\sqrt{-g} g^{nk} \partial_{k} {\hat X}^{\underline{m}}
\right)\right]
$$
$$
+ \int\limits^{}_{\partial {\cal M}^{2}} \delta
  P_{1\underline{m}} \left[ \hat{X}^{\underline{m}}(\tilde{\xi } (\tau )) -
\tilde{X}^{\underline{m}} (\hat{\zeta} (\tau ) )\right] -
$$
$$
-\int\limits^{}_{{\cal M}^{p+1}} d^{p+1} \zeta
\left[\partial_{\tilde{m}} \left(
\sqrt{|G|}
{G}^{-1 [\tilde{m}\tilde{n}] }\right)
- j^{\tilde{n}} \right] \delta A_{\tilde{n}} (\zeta )
$$
$$
- \int_{{\cal M}^{p+1}}  \left[d^{p+1} \zeta \partial_{\tilde{m}}\left(
\sqrt{|G|} G^{-1 (\tilde{m}\tilde{n})}
\partial_{\tilde n}\tilde{X}^{\underline{m}} \right)
+ j_{p} \wedge P_1^{\underline{m}} \right] \delta  \tilde{X}_{\underline{m}}.
$$

Then the equations of motion become evident.
They are

\begin{itemize}
\item
The Born--Infeld equation
(with $G_{mn}$ defined by (\ref{GgF}))
\be\label{BI}
\partial_m \left(\sqrt{|G|}
{G}^{-1 [\tilde{m}\tilde{n}] }\right)
= j^{\tilde{n}} \equiv
 \int_{\partial {\cal M}^2} d\hat{\zeta}^{\tilde{n}}(\tau )
\delta^{p+1}\left({\zeta}-\hat{\zeta}(\tau )\right).
\ee
It includes the definite source localized on the intersection
(and thus outside the intersection the equation is free).
\item
Equations for the Dp-brane coordinate functions
(with $G_{mn}$ defined by Eq. (\ref{GgF}))
\footnote{
To arrive at the expression for the r.h.s. one should perform some formal
extension of the property (\ref{j31}), as the form $P_1= d\tau {\cal P}(\tau)$
has  not been defined as a pull--back of any 1-form on ${\cal M}^{p+1}$.
Let us assume that such a form
$
\tilde{P}_{1\underline{m}} = d\zeta^{\tilde{m}}
 \tilde{{\cal P}}_{\underline{m}}$
is introduced.
Then, by definition,
$
\tilde{P}_{1\underline{m}} (\hat{\zeta}(\tau ))$ $ \equiv $ $
d\tau \partial_\tau \hat{\zeta}^{\tilde{m}} \tilde{{\cal P}}_{\underline{m}}
(\hat{\zeta}(\tau ))
= d\tau {{\cal P}}_{\underline{m}}(\tau)
$
and one finds
$
\int_{{\cal M}^{p+1}} j_p \wedge  \tilde{P}_{1\underline{m}} =
\int_{\partial {\cal M}^2} d\tau {\cal P}_{\underline{m}} (\tau).
\delta^{p+1} (\zeta - \hat{\zeta}(\tau)$.
}
\be\label{vX}
\partial_m \left(\sqrt{|G|}
{G}^{-1 (\tilde{m}\tilde{n}) }
\partial_{\tilde{n}} \tilde{X}^{\underline{m}}
\right)
= - \int_{\partial {\cal M}^2} d\tau {\cal P}^{\underline{m}} (\tau)
\delta^{p+1}\left({\zeta}-\hat{\zeta}(\tau )\right)
\ee

\item The equation for the string coordinate functions
(with $g_{mn}(\xi )$ defined by (\ref{imstr}) or, equivalently, by
(\ref{vg}) and the Weyl symmetry (\ref{Wg}))
\be\label{vXsi}
d^2 \xi \partial_{m} \Big( \sqrt{-g} g^{mn}(\xi ) \partial_{n}
{\hat X}^{\underline{m}}(\xi ) \Big)
= j_1 \wedge  \left( P_1^{\underline{m}} + d \xi^m \varepsilon_{mn}
\sqrt{-g} g^{nk} \partial_{k} {\hat X}^{\underline{m}}
\right).
\ee
\end{itemize}

\section{Reparametrization symmetry of  the coupled system}

\bigskip

The variations of the action (\ref{Sg}) with respect to the worldvolume
and worldsheet embedding coordinate functions $\delta \hat{\zeta} (\tau
)$ (\ref{bD}) and ${\tilde \xi}^m (\tau) $   (\ref{bembb}) is
 \begin{equation}\label{varSzx}
\delta_{\tilde{\xi} , \hat{\zeta} } S=
\int_{\partial {\cal M}^{2}} \delta \hat{\zeta}^{\tilde{m}} (\tau )
\left[ i_{\tilde{m}}{\cal F} - P_{1\underline{m}}\partial_{\tilde{m}}
\tilde{X}^{\underline{m}}\right] +
\end{equation}
$$
+ \int_{\partial {\cal M}^{2}} \delta \tilde{\xi}^{{m}} (\tau )
\left[ P_{1\underline{m}} \partial_{{m}}
\hat{X}^{\underline{m}} -
d \tilde{\xi}^p(\tau)  \varepsilon_{pn}
\sqrt{-g} g^{nk} \partial_{k} {\hat X}^{\underline{m}}
\partial_m {\hat X}_{\underline{m}}
\right].
$$

The corresponding equations of motion
\be\label{vzeta}
i_{\tilde{m}} {\cal F} \vert_{\partial {\cal M}^{2}} =
P_{1\underline{m}}
\partial_{\tilde{m}} \tilde{X}^{\underline{m}} (\zeta (\tau))
\ee
\be\label{vxi}
d \hat{\xi}^p(\tau)  \varepsilon_{pn}
\sqrt{-g} g^{nk} \partial_{k} {\hat X}^{\underline{m}}
\partial_m {\hat X}_{\underline{m}}
=
P_{1\underline{m}} \partial_{{m}}
\hat{X}^{\underline{m}}
\ee
are dependent.
%These are  the Noether identities
%whish reflect the
%presence of counterparts of both stringy and Dp--brane reparametrization
%symmetry in the action for the coupled system (\ref{Sg}).

To prove the dependence of Eq. (\ref{vxi}) one should
contract Eq. (\ref{vXsi}) with $\partial_l \hat{X}_{\underline{m}} (\xi)$
on the target space indices. Then, writing the left hand part
as
\be\label{tra}
\partial_{m} \Big( \sqrt{-g} g^{mn}(\xi ) \partial_{n}
{\hat X}^{\underline{m}}(\xi ) \Big)
\partial_l{\hat X}^{\underline{m}}(\xi )
=
\ee
$$
= \partial_{m} \Big( \sqrt{-g} g^{mn}(\xi ) \partial_{n}
{\hat X}^{\underline{m}}(\xi )
\Big)
\partial_l{\hat X}^{\underline{m}}(\xi )
\Big) -
\Big( \sqrt{-g} g^{mn}(\xi ) \partial_{n}
{\hat X}^{\underline{m}}(\xi )
\Big)
\partial_l
\partial_{m}
{\hat X}^{\underline{m}}(\xi )
$$
and using Eq. (\ref{vg}) (which remains the same for the coupled
system), one obtains (in arbitrary space-time dimension D)
\be\label{vXst}
d^2 \xi
\sqrt{-g}
{D-2 \over 4}
\partial_{l} \Big( \sqrt{-g} g^{mn}(\xi )
\partial_{m}{\hat X}^{\underline{m}}(\xi )
\partial_{n}{\hat X}^{\underline{m}}(\xi )
\Big)
=
\ee
$$
- j_1 \wedge  \left( P_1^{\underline{m}} + d \xi^m \varepsilon_{mn}
\sqrt{-g} g^{nk} \partial_{k} {\hat X}^{\underline{m}}
\right).
$$
However, using the Weyl symmetry one can fix the gauge (\ref{Wgg}),
where the l.h.s.\ of Eq.\ (\ref{vXst}) vanishes, and we arrive at
 (\ref{vxi}).

To prove the dependence of Eq. (\ref{vzeta}) one should
consider the longitudinal part of Eq. (\ref{vX}), i.e the result of the
contraction  of Eq. (\ref{vX})  with
$\partial_{\tilde{m}} \tilde{X}^{\underline{m}}$
on the target space
indices
\be\label{vXt}
\partial_m \left(\sqrt{|G|}
{G}^{-1 (\tilde{m}\tilde{n}) }
\partial_{\tilde{n}} \tilde{X}^{\underline{m}}
\right)
\partial_{\tilde{l}} \tilde{X}_{\underline{m}}
=
- p^{\underline{m}}
\partial_{\tilde{l}} \tilde{X}_{\underline{m}}.
\ee
Here we,  for shortness,
introduced the notation
$ \int_{\partial {\cal M}^2} d\tau {\cal P}^{\underline{m}} (\tau)
\delta^{p+1}\left({\zeta}-\hat{\zeta}(\tau )\right)
$
$
=-p^{\underline{m}}.
$
Dealing with the l.h.s.\ of Eq.\ (\ref{vXt}) in the same way as in
(\ref{tra}) and using the identities
$$
 \delta^{\tilde{m}}_{~\tilde{n}} =
 G^{-1 (\tilde{m}\tilde{k})}
 \partial_{\tilde{k}}\tilde{X}^{\underline{m}}
 \partial_{\tilde{n}}\tilde{X}_{\underline{m}}
 +
 G^{-1 [\tilde{m}\tilde{k}]}
 F_ {\tilde{k}\tilde{n}}, \qquad
$$
$$ G^{-1 [\tilde{m}\tilde{k}]}
 \partial_{\tilde{k}}\tilde{X}^{\underline{m}}
 \partial_{\tilde{n}}\tilde{X}_{\underline{m}}
= -
 G^{-1 (\tilde{m}\tilde{k})}
 F_ {\tilde{k}\tilde{n}}, \qquad
$$
 (which follow from $G^{-1 \tilde{m}\tilde{k}} G_{\tilde{k}\tilde{n}} =
 \delta^{\tilde{m}} _{~\tilde{n}}$) one arrives at
\be\label{vXt2}
\partial_m \left(\sqrt{|G|}
{G}^{-1 (\tilde{m}\tilde{n}) }
\partial_{\tilde{n}} \tilde{X}^{\underline{m}}
\right)
\partial_{\tilde{l}} \tilde{X}_{\underline{m}}
= -
\partial_m \left(\sqrt{|G|}
{G}^{-1 [\tilde{m}\tilde{k}] }
\right)
F_{\tilde{k}\tilde{n}}.
\ee
Substituting Eqs. (\ref{vXt}) and (\ref{BI}) we obtain from Eq.
(\ref{vXt2})
\be\label{vXt3}
    \int_{\partial {\cal M}^2} \left(- i_{\tilde{m}} F_{2} +
P_{1\underline{m}}
\partial_{\tilde{m}} \tilde{X}^{\underline{m}} (\zeta (\tau))
 \right)
\delta^{p+1} \left( \zeta - {\hat \zeta} (\tau )\right)~= 0.
\ee
Eq. (\ref{vXt3}) is equivalent to  (\ref{vzeta}).

The dependence of the equations
(\ref{vzeta}), (\ref{vxi}) should be regarded as
Noether identities which reflect some gauge symmetries
 of the
action.

Reversing the arguments, we can consider
(\ref{vzeta}), (\ref{vxi})  as independent equations, but
conclude that the longitudinal parts of Eqs. (\ref{vX}), (\ref{vXsi}) are
dependent as they are in the case of the free string and
Dp-brane. From such point of view we can state that
the dependence of Eq. (\ref{vxi}) and Eq. (\ref{vzeta}) reflects
 the reparametrization symmetries of the string worldsheet and the Dp-brane
 worldvolume
 which, hence, survive when coupling is 'switched on'.

\section{Concluding remarks}

Thus, applying the method of Ref. \cite{BK} we obtained the
 equations of motion  (\ref{BI}) -(\ref{vXsi}), (\ref{id})
for the coupled system of an open bosonic string and
a dynamical bosonic Dp--brane where the string endpoints live.
In such a way we illustrated the key issues of the approach
\cite{BK,BKD9} and we are able to comment on its relation
to the standard
description of Dp--branes as rigid planes \cite{Dbranes}.

The investigation of Eqs.
(\ref{BI}) -(\ref{vXsi}), (\ref{id})
and their  supersymmetric
generalizations will be the subject of the forthcoming paper.
Note that such a study
simplifies essentially when the Lorentz--harmonic
formulations of the (super--)string \cite{BZ} and (super--)Dp--brane actions
\cite{bst} are used instead of (\ref{Sbstr}), (\ref{Dp}).

\section*{Acknowledgments}
The authors are grateful to D. Sorokin and M. Tonin for interest in
this work and many useful conversations.
One of the authors (I.B.) thanks
Padova Section of INFN for hospitality in Padova, where a part of this work
has been done.
A partial support from the INTAS Grant {\bf 96-308} and from the
Ukrainian GKNT Grant {\bf 2.5.1/52} is acknowledged.

%\newpage

\end{document}